\documentclass[12pt]{article}

\usepackage{amsmath,amssymb}
\usepackage{graphicx}

\usepackage{cite}
\usepackage{hyperref}

\usepackage[margin=1in]{geometry}

\usepackage{setspace} 
\makeatletter
\renewcommand{\@biblabel}[1]{\quad#1.}
\makeatother

\begin{document}
\begin{flushleft}
{\Large
\textbf{Inference for changes in biodiversity}
}
\\
Amy Willis$^{1,\ast}$, 
John Bunge$^{1}$, 
Thea Whitman$^{2,3}$,
\\
\bf{1} Department of Statistical Science, Cornell University, Ithaca, NY, USA \\
\bf{2} Crop and Soil Sciences, Cornell University, Ithaca, NY, USA
\\
\bf{3} Department of Environmental Science, Policy and Management, University of California, Berkeley, CA, USA
\\
$\ast$ E-mail: adw96@cornell.edu
\end{flushleft}

\section*{Abstract}

We wish to formally test for changes in the taxonomic diversity of a community, especially in the presence of high latent diversity. Drawing on the meta-analysis literature, we construct a model for diversity that accounts for covariate effects as well as sampling variability. This permits inference for changes in richness with covariates and also a test for homogeneity. We argue that we can use the principles of shrinkage estimation to improve richness estimation in this nonstandard context, which is especially important given the high variance of richness estimators and the increasing abundance of community composition data. We demonstrate the methodology under simulation, in a gut microbiome study (testing for a decrease in richness with antibiotics), and in a soil microbiome study (testing for homogeneity of replicates). We believe that this is the first formal procedure for analyzing changes in species richness.


\section*{Introduction}

The taxonomic diversity of a biological population is commonly used as a marker for ecosystem health. However, taxonomic diversity, or {\it species richness}, is sensitive to changes in the ecosystem, for example, temperature, time, biogeochemical conditions, and anthropogenic factors. Understanding the effects of mechanisms that may incite or accelerate changes in diversity is crucial to sustaining ecosystem balance. For this reason, many micro- and macroecologists are interested in testing for statistically significant changes in diversity in response to one or more covariates. Such inference informs conclusions about sustaining the health of the natural ecosystem, whether terrestrial, aquatic, or even within the human body.

Modelling diversity is more complex than may first appear. This is because sampling from an environment, for instance, the microbial population in a patch of soil, or butterflies in a rainforest, is almost always inexhaustive. As a result, any diversity analysis based solely on the observed sample fails to account for the unobserved diversity -- the taxa in the population that eluded the sample. In ecosystems where the total diversity is low relative to observed diversity, unobserved species may not greatly threaten the validity of conclusions drawn that only consider observed diversity. However, in high-diversity ecosystems, unobserved species may comprise the bulk of the total number of species, and only a small fraction of the total number of species may be believed to be observed in the sample. The latter case is especially prevalent in skin, water and soil microbiome studies, because these ecosystems are characterised by large numbers of very rare species. 

In this paper we propose a method for modelling species richness that considers both the observed {\it and} unobserved members of the population. This allows us to draw conclusions about the population under study, rather than merely conclusions about the samples that were observed. The number of unobserved members of the population is unknown and must be estimated, and the error in estimation must be accounted for in the diversity model. The resulting procedure permits formal statistical inference about which covariates affect total diversity. In addition, the procedure admits a natural test for homogeneity of taxonomic richness of samples. Homogeneity is of particular importance because the experimentalist is often interested in the consistency of results across biological replicates.

A brief overview of the paper is as follows: we begin by discussing some recent published work in the ecological literature which examined diversity as a function of covariates, and discuss the techniques used by these authors in making their conclusions. We argue that these methods do not account for unobserved diversity. We then introduce our model for species richness and discuss parameter estimation. We show some simulation results justifying model assumptions and conclude with an examination of the dataset of Dethlefsen {\it et. al}, where we conclude that an antibiotic significantly decreases gut microflora diversity; and the dataset of Whitman {\it et. al}, where we conclude homogeneity of biological replicates in the soil microbiome.

\section*{Existing methods for diversity quantification}

Interest in modelling changes in species richness spans a broad array of ecological fields. The effect of climate change on species richness has been of particular recent interest \cite{globalwarming1,globalwarming2,globalwarming3}, as has the link between human microbiome diversity and health \cite{gutmicrobiome, vaginalmicrobiome}. A survey of recent papers examining changes in diversity revealed that the most common technique used to conclude that diversity changes with a covariate is to fit a linear or nonlinear regression model for observed richness as a function of the covariate \cite{globalwarming1, globalwarming3,lhkf,gutmicrobiome,rodentparasite,vaginalmicrobiome}. If the regression coefficient of the covariate is statistically significant, then a relationship is concluded. Usually such conclusions are accompanied with a plot of observed diversity against the covariate and comments justifying the choice of model; for example, linear versus nonlinear. 

Another common approach to quantifying diversity is to model a sample diversity index, the most common of which are the Simpson index and the Shannon index. While use of diversity indices is very popular in the ecological literature, there are numerous problems associated with the manner in which they are employed, the most serious of which are estimation and confounding. For example, the population Simpson diversity is $$H_{Simpson} = \sum_{i=1}^C p_i^2$$ where $p_i$ is the true probability of observing the $i$th taxa and $C$ is the true species richness. However, since usually only a subset (sample) of the population is observed, the probabilities $p_i$ are unknown and must be estimated. While $\hat{p}_i = n_i/n$ (where $n_i$ is the number of times the $i$th taxa is observed in a sample of size $n$) is a statistically optimal estimator of $p_i$ under the assumption of sampling independence (it has minimum variance out of all unbiased estimators), it is not the case that the plug-in estimator of $H_{Simpson}$ is statistically optimal, where the plug-in estimator is defined to be $$\hat{H}_{Simpson, \text{plug-in}} = \sum_{i=1}^c \hat{p}_i^2,$$ where is $c$ the observed species richness. Indeed, the minimum variance unbiased estimator was derived only very recently \cite{zhzh}. The other issue with diversity indices is their interpretation. Smaller values of the Simpson index are associated with a larger number of classes in the population, and also with more inequality in their proportions. Larger values of the Shannon index are associated with a larger number of classes, but with more equality in their proportions. For this reason, it is not possible to conclude whether a population is very diverse, or alternatively very even, based solely on the information provided by the value of a diversity index. In this way, diversity indices confound the effects of evenness and richness, resulting in limited interpretation of their values as summaries of the sample. For these reasons we discourage the use of diversity indices in the ecological literature and focus discussion of this manuscript on modelling total species richness. 

We now propose a technique for modelling species richness in a way that accounts for the estimation of unobserved diversity. We note here that if a practitioner was interested in modelling diversity indices, they could proceed in the same way -- that is, by taking account of the error in their estimation using the standard errors derived in \cite{zhzh} and utilizing them as in the introduced model for species richness.

\section*{A model for total species richness}

Suppose we are interested in comparing the total species richness across $m$ populations.  Denote the total richness in the $i$th population, observed and unobserved, by $C_i, i=1,\ldots,m$.  Also associated with each population is a set of $p$ covariates, often called ``metadata''. We assume that species richness a function of the covariates, but also a function of pure random variation, so that
\[
C_i = \beta_0 + \beta_1 x_{i,1} + \ldots + \beta_p x_{i,p} + u_i,
\]
where $x_{i,j}$ is the value of the $j$th covariate for the $i$th population, $\beta_j$ is its coefficient ($j=1,\ldots,p$), and $u_i$ is a random variable representing the variation in richness not attributable to the covariates.  In matrix terms,
\[
\vec{C} = \mathbf{X} \vec{\beta} + \vec{U},
\]
where $\vec{C} = [C_1 \cdots C_m]^T, \vec{\beta} = [\beta_0 \cdots \beta_p]^T, \vec{U} = [u_1 \cdots u_m]^T$, and $\mathbf{X} = [\vec{1} \mbox{ } \vec{x}_1 \cdots \vec{x}_p]$ with $\vec{x}_j = [x_{1,j} \cdots x_{m,j}]^T$.   We make the assumption that $u_i,\ldots,u_m$ are independent, identically distributed normal random variables with common variance $\sigma^2_u$: $\vec{U} \sim N(\vec{0},\sigma^2_u I_m)$. We assume that we have complete information about the covariates associated with each population, that is, the ``design matrix'' $\mathbf{X}$ is known. Note that while the model is technically linear, nonlinear terms may be incorporated through $\mathbf{X}$ as usual in a regression analysis.

Suppose the goal of the experiment is to investigate which covariates do and do not alter total species richness, or equivalently, which elements of $\beta$ are equal to zero. This type of analysis is of very broad interest and answering it comprised some component of each \cite{globalwarming1,globalwarming2,globalwarming3,
gutmicrobiome,vaginalmicrobiome,lhkf,rodentparasite,yeasts,
nbacteria, aranda}. In order to answer this question, we take a sample of individuals from each of the $m$ populations under study. We do {\em not} assume equal sample sizes (sampling depth), or that every taxon in each population was observed. Because we do not assume that every taxon in each population was observed, we do not know $C_i$ exactly for any $i$: the total species richness is unknown for each of the populations under study. Consequently our inference about $\vec{\beta}$ requires accounting for error in estimation of $\vec{C}$. 

Based on each of our samples, we estimate $C_i$ by $\hat{C}_i$ with standard error $\hat{\sigma}_i$. A large number of estimators for species richness have been developed since the introduction of the problem into the statistical literature in 1943 \cite{fcwm}. For a recent review, see \cite{bwwa}, and for a diversity estimator developed for the high-diversity microbial setting see \cite{rbmeunc,breakaway}. For estimators based on maximum likelihood or nonlinear regression, asymptotic theory ensures the asymptotic normality of diversity estimates under the assumption of correct model specification (see \cite{rbmeunc} and references therein; we also investigate this with simulations below).  We therefore assume that, conditional on the value of $C_i$, the estimate $\hat{C}_i$ is normally distributed around $C_i$ with standard deviation $\sigma_i$, that is,
\[
\hat{C}_i|C_i = C_i + \epsilon_i,
\]
where $\epsilon_i \sim \mathcal{N}(0,\sigma^2_i), i=1,\ldots,m$.  Unconditionally we then have the final model 
\[
\hat{C}_i = \beta_0 + \beta_1 x_{i,1} + \ldots + \beta_p x_{i,p} + u_i + \epsilon_i,
\]
or in matrix terms
\begin{equation} \label{mainmodel}
\vec{\hat{C}} = \mathbf{X} \vec{\beta} + \vec{U} + \vec{\epsilon},
\end{equation}
where
\[
\vec{\epsilon} = \left[ \begin{array}{c} \epsilon_1 \\ \vdots \\ \epsilon_m \end{array} \right] \sim \mathcal{N}\left(\vec{0},\left[ \begin{array}{cccc} \sigma_1^2 & 0 & \cdots & 0 \\ 0 & \sigma_2^2 & \cdots & 0 \\ \vdots & \vdots & \ddots & \vdots \\ 0 & 0 & \cdots & \sigma_m^2 \end{array} \right] \right) ,
\]
and $\vec{U} \sim \mathcal{N}(0,\sigma^2_u I_m)$. Since the only available information on $\sigma_i$ is the standard error $\hat{\sigma}_i$ we substitute the latter for the former and henceforth refer only to $\sigma_i$.  We evaluate the effect of this substitution via simulation below.

It is important to note that the stochastic nature of the estimated total diversity arises both from $\vec{U}$ and $\vec{\epsilon}$, that is, through the inherent random variation of the $C_i$ around $\mathbf{X}\vec{\beta}$ and through the random variation of $\hat{C}_i$ around $C_i$.   Procedures that model the observed diversity $n_i$ as a linear function of the covariates effectively set $\hat{C}_i = n_i$ but treat $\sigma^2_i=0$, thus treating the sample as the population and unobserved diversity as null.

Given model (\ref{mainmodel}) there are two main hypotheses of interest.  The first is $H_0: \sigma^2_u = 0$; that is, the variation in the true species richnesses across the $m$ populations is wholly attributable to the covariates $x_1, \ldots, x_p$ with no unexplained random variation. This hypothesis is often referred to as that of {\it homogeneity}. The alternative hypothesis of  {\it heterogeneity}, $H_A: \sigma^2_u>0$, supposes that there is more variability in the diversity estimates than can be explained by sampling-based variation in the estimates alone, and that some other mechanism (which we ascribe to the random variables $u_1, \ldots, u_m$) contributes to the observed behavior of the estimated species richnesses.  Rejection of $H_0$ does not imply that the diversity model is misspecified, merely that there is some factor which alters the true total richnesses that is not accounted for in the model. For instance, a predictor variable that affects species richness but is absent from the model may result in rejection of $H_0$ even if the populations are homogeneous. However, it is also possible to include all informative predictors of richness and still have true heterogeneity: this conclusion implies that the species richness of the populations is distinct without further cause.

The second main hypothesis of interest is $H_0: \beta_1 = \ldots = \beta_p = 0$, or alternatively, that none of covariates explain the variation in richness across populations.  The alternative hypothesis $H_A$ is then that at least one of the covariates affects richness.  If $H_0$ is rejected then interest focuses on the covariates $x_j$ that do influence richness: which $\beta_j$ are nonzero and what are their magnitudes? The relevant null hypothesis for the case of one variable is then $H_0: \beta_j=0$. Note that the usual regression interpretation of the coefficients applies and that $\beta_j$ is the expected increase in the true diversity of any of the $i$ populations for a one unit increase in $x_{i,j}$. 


It remains to discuss estimation of the model parameters $\vec{\beta}$ and $\sigma^2_u$, and implementation of the stated hypothesis tests. The log-likelihood of our model is $$l(\vec{\beta},\sigma^2_u | \mathbf{X},\hat{C}_1,\ldots,\hat{C}_m, \sigma^2_1,\ldots,\sigma^2_m) = -\frac{1}{2} \sum_{i=1}^m \left[ \ln \left(\sigma^2_u + \sigma^2_i\right) + \frac{(\hat{C}_i-\vec{x}_i^T\vec{\beta})^2}{\sigma_u^2 + \sigma^2_i} \right].$$
Maximum likelihood (ML) is a natural choice of parameter estimation technique due to its many asymptotic and finite sample optimality properties in standard settings \cite{casellaberger,godambe}. However, in this application the choice to use ML is non-trivial because of the {\it boundary problem}: $\sigma_u^2 \geq 0$. This problem was studied by Crainiceanu and Ruppert \cite{ccdr}, who demonstrate the failure of the usual likelihood ratio test asymptotics when testing $\sigma^2_u=0$ against $\sigma^2_u >0$.

Fortunately, we can exploit the well-developed literature on meta-analysis to resolve these difficulties. Meta-analyses arise in many social and health sciences where a researcher wishes to pool a number of different (and often disagreeing) studies to determine the presence of an overall effect. Each richness estimate fulfills the role of a study's effect estimate, the standard error of the richness estimate fulfills the role of the standard error of the effect estimate, and the $m$ populations of interest reflect $m$ different studies to be pooled. A comprehensive treatment of meta-analyses is given by Demidenko \cite{demi}, who discusses both restricted maximum likelihood algorithms and also the best choice of hypothesis test in this nonstandard boundary case. We note also that in species richness comparison, as with meta-analyses, we only know the standard error in the estimates $\hat{\sigma}_i$ and not the true standard deviations $\sigma_i$. For this reason we base our choice of asymptotics on the results of \cite{demi} rather than those of \cite{ccdr}. Following the theory laid out by \cite{demi}, our restricted ML procedure maximizes $$l_R(\vec{\beta},\sigma^2_u ) = -\frac{1}{2} \left\{ \sum_{i=1}^m \left[ \ln \left(\sigma^2_u + \sigma^2_i\right) + \frac{(\hat{C}_i-\vec{x}_i^T\vec{\beta})^2}{\sigma_u^2 + \sigma^2_i} \right] +\ln \left| \sum_{i=1}^m\frac{\vec{x}_i\vec{x}_i^T}{\sigma^2_u+\sigma^2_j}\right| \right\},$$ and we denote the maximizing values by $\hat{{\beta}}$ and $\hat{\sigma}^2_u$. Unfortunately there does not exist closed form expressions for $\hat{{\beta}}$ and $\hat{\sigma}^2_u$ but we have the following update procedures:
\begin{align}
\hat{{\beta}}_{s+1} &=  \left( \sum_{i=1}^m \frac{\vec{x_i}\vec{x_i}^T}{\hat{\sigma}^2_{u,s} + \sigma^2_i}\right)^{-1} \sum_{i=1}^m \frac{\vec{x}_i y_i}{\hat{\sigma}_{u,s}^2 + \sigma^2_i}, \notag \\
\hat{\sigma}^2_{u,s+1} &= \left( \sum_{i=1}^m \frac{1}{\hat{\sigma}^2_{u,s} + \sigma^2_i}\right)^{-1} \left[ \sum_{i=1}^m \frac{(\hat{C}_i - x_i^T\hat{\beta}_s)^2-\sigma^2_i}{(\hat{\sigma}^2_{u,s} + \sigma_i^2)^2}+G(\hat{\sigma}_{u,s}^2)\right], \notag \\
G(\sigma^2) &= tr \left( \left( \sum_{i=1}^m \frac{\vec{x}_i \vec{x}_i^T}{\sigma^2 + \sigma^2_i}\right)^{-1} \left( \sum_{i=1}^m \frac{\vec{x}_i \vec{x}_i^T}{(\sigma^2 + \sigma^2_i)^2} \right) \right) \notag
\end{align}
where in the above and henceforth we assume that the vector $\beta$ includes the intercept coefficient and accordingly that $\vec{x}_{1,i}=1$ for $i=1,\ldots,m$ (we retain that $p$ represents the number of non-intercept predictors). Robust choices of starting values $\hat{\beta}_0$ and $\hat{\sigma}^2_0$ are also given in \cite{demi}. Our investigations lead us to conclude that the least squares estimate of $\beta$ (obtained by regressing the covariates on the observed diversity) is a reasonable value for $\hat{\beta}_0,$ and the empirical variance in the estimates $\hat{C}_i$ is a reasonable value for $\hat{\sigma}^2_0$.

Without boundary complications, hypothesis testing for $\beta$ falls in the standard Wald-type framework. Inverting second derivatives of the restricted log-likelihood gives the variance estimate
$$\hat{\text{Var}}(\hat{\beta}) = (\mathbf{X}^T \hat{\mathbf{W}}^{-1}\mathbf{X})^{-1},$$ 
where $\hat{\mathbf{W}} = \text{diag} (\hat{\sigma}_1^2 + \hat{\sigma}_u^2,\ldots,\hat{\sigma}_m^2 + \hat{\sigma}_u^2),$
which we use to make marginal inference about the effect of each predictor on species richness via the test statistic $\frac{\hat{\beta}_i}{\sqrt{\hat{[\text{Var}}(\hat{\beta})]_{ii}}}$, which is distributed approximately $\mathcal{N}(0,1)$. The global test of $H_0: \beta_1 = \ldots = \beta_p=0$ has test statistic 
$$\hat{\beta}_{-1}^T \mathbf{X}^T_{-1} [\hat{\mathbf{W}}^{-1}]_{-1}\mathbf{X}_{-1}\hat{\beta}_{-1},$$
which is distributed asymptotically according to a $\chi^2_p$ distribution (the subscript denotes the omission of the intercept term). Finally, we define our $Q$-statistic as $$Q = \sum_{i=1}^m \frac{(\hat{C}_i- \vec{x}_i^T\hat{\beta})^2}{\hat{\sigma}_i^2}.$$ Under the null hypothesis of homogeneity, $Q$ follows a $\chi^2$ distribution with $m-p-1$ degrees of freedom.

\subsection*{Improving richness estimation}

While most studies are interested in modelling diversity with covariates, it may be the case that the practitioner is interested in species richness of itself. As discussed previously, species richness estimation is known to be a difficult problem in statistics: most models that offer good fits to frequency count data supply large standard errors, which can be traded off at the expense of highly parametrized models \cite{rbmeunc, catchall}. Species richness estimation has hitherto been based on the outcomes of a single experiment in which one obtains the number of species observed once (the {\it singletons}), the number of species observed twice (the {\it doubletons}), and so forth, and uses this information to predict the number of species that were not observed. This prediction of unobserved richness is added to the observed richness for an estimate of the total richness. In the setting discussed in this paper, information is available about multiple experiments of a similar nature: for instance, samples of the microbial composition of the gut of a number of study participants. It is very likely that the gut microbiomes of the participants possess similar structural qualities. Therefore, given the high-variance nature of richness estimation of a single sample, it is desirable to take advantage of common features across multiple samples to reduce variability in richness estimation. This concept is often referred to as {\it shrinkage estimation} in the statistical literature. 

Recall that our model for estimated species richness is $$\vec{\hat{C}} = \mathbf{X}\vec{\beta} + \vec{U} + \vec{\epsilon},$$ where $\vec{U} \sim \mathcal{N}(0,\sigma^2 I_m)$ is the random effect term that reflects the random variation in true species that is not attributable to the covariates. Thus, each $u_i$ constitutes a realisation of random variable, but because we only have access to $\hat{C}_i$ and not to $C_i$, we do not know the realisation of the random variable $U_i$ and may only predict it. If we let $\hat{U}^*$ denote our prediction of the random vector $U$, the estimate of the total richnesses given by $$\hat{C}^* = \mathbf{X}\hat{\vec{\beta}} + \hat{U}^*$$ provides a lower variance estimate than that given by $\hat{C}.$ We argue that if estimates of total species richness are of interest, combining an estimate of $\beta$ with a prediction of $U$ permits strength in prediction to be shared across multiple samples, resulting in lower variance estimates of richness compared to those based simply on frequency count data.

The question remains of how to best predict the random variable $U$. The most common approach to prediction of random effects is known as {\it best linear unbiased prediction}, or BLUP \cite{henderson, blupreview}. Making the appropriate substitutions to the methodology described in \cite{sas}, we have that $$\hat{U}^*_i = \frac{\hat{\sigma}_u^2}{\hat{\sigma}_i^2+\hat{\sigma}^2_u} (\hat{C}_i - \vec{x}_i^T\hat{\beta}).$$ The estimated variance of the predictions is given in \cite[p751--752]{sas} and is a function of the variance of $\hat{\beta}$, the variance of $\hat{\sigma}^2_u$, and the covariance between $\hat{\beta}$ and $\hat{\sigma}^2_u$. Note that this estimate of the variance does not account for the estimated nature of $\hat{\sigma}^2_u$ and $\hat{\sigma}^2_i$ and so may underestimate the true sampling variability when the design is unbalanced and no replicates are available. Investigation of the extent of this variance underestimation and appropriate corrections is an ongoing topic of investigation by the authors.

\subsection*{Model selection and diagnostics}
The methodology described above is sensitive to the model design $\mathbf{X}$, and method of estimating $C$. Perusal of the richness estimates, and more importantly, their standard errors, is essential to ensure that the model is not overfit (with respect to predictors) and heterogeneity is not falsely concluded. One exploratory approach to diagnosing for possible outliers is to plot the estimated richness with error bars at $\pm 2$ standard errors. For an example, see Figures \ref{fig3} and \ref{fig3a}. This technique derives from, and is limited by, the assumption that the estimated richness are normally distributed around the true estimated richnesses with standard deviation equal to the estimated standard error. The visual diagnostic for an outlier is a tight interval (small estimated error), especially one centered far from the overall mean. While it is tempting to notice the large intervals in this type of plot, in fact these types of points do not exert a large influence on the model because most of the variability is captured in the large ``local error'' ($\sigma_i^2$) rather than affecting the estimate of the ``global error'' ($\sigma_u^2$). 


While visual diagnostics of this nature can assist with model selection and formulating appropriate hypotheses, graphical procedures such as this one suffer from the problem of simultaneous inference. For this reason, ``testing'' multiple 95\% confidence intervals for overlap has an exaggerated probability of committing a Type 1 error. For this reason we advocate the mixed model procedure described above, which does not require multiple testing corrections. We proceed to demonstrate the method with some examples.

\section*{Results}


\subsection*{Simulations}

\begin{center}
\begin{figure}
\includegraphics[trim= 0mm 0 0 0, scale=0.5]{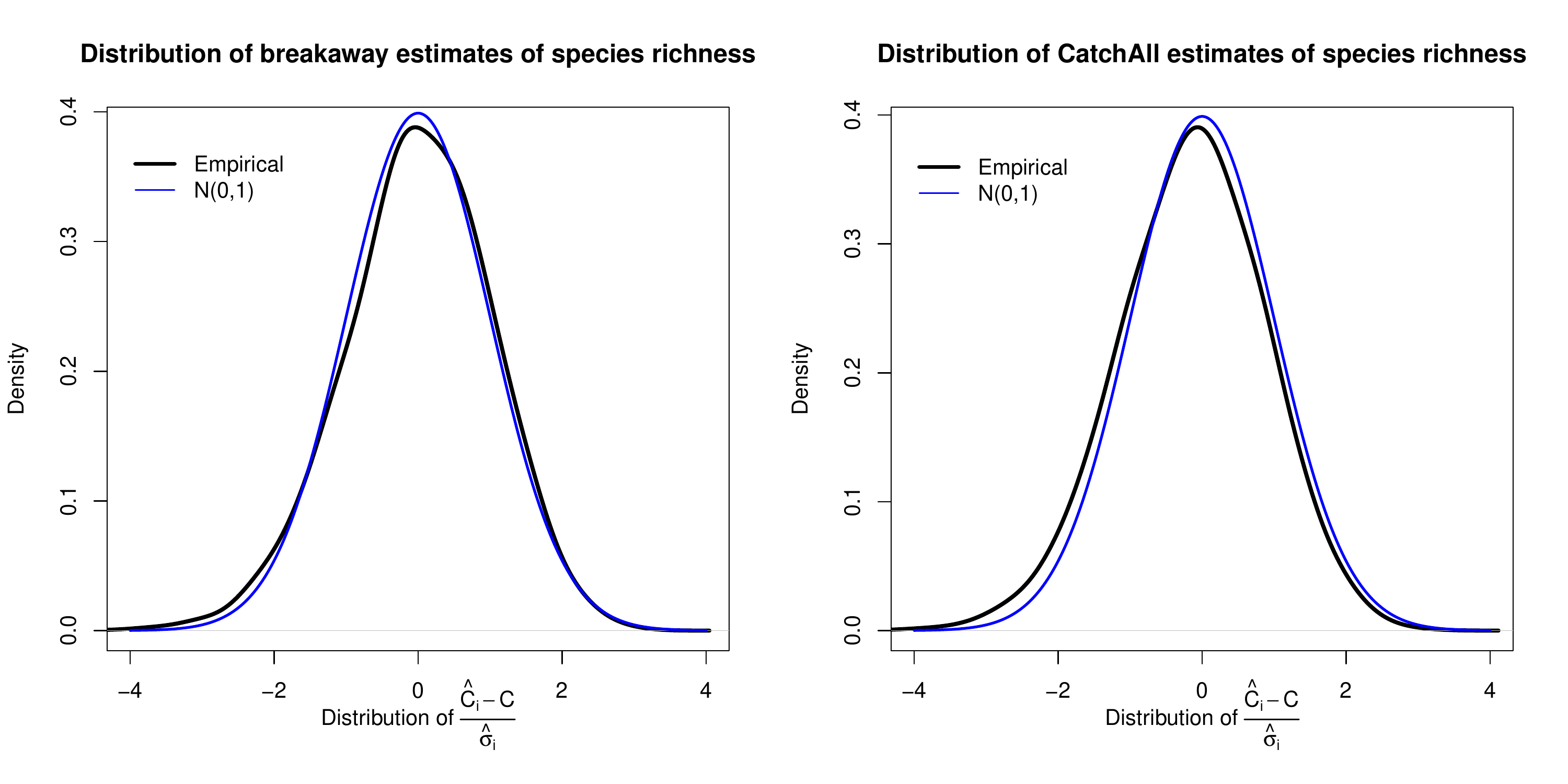}
\caption{Richness estimates derived from software breakaway (left) and CatchAll (right) appear to be approximately normally distributed with correct standard error in the negative binomial counts case.}
\label{fig1}  
\end{figure}
\end{center}

\begin{center}
\begin{figure}
\includegraphics[trim= 0mm 0 0 0, scale=0.5]{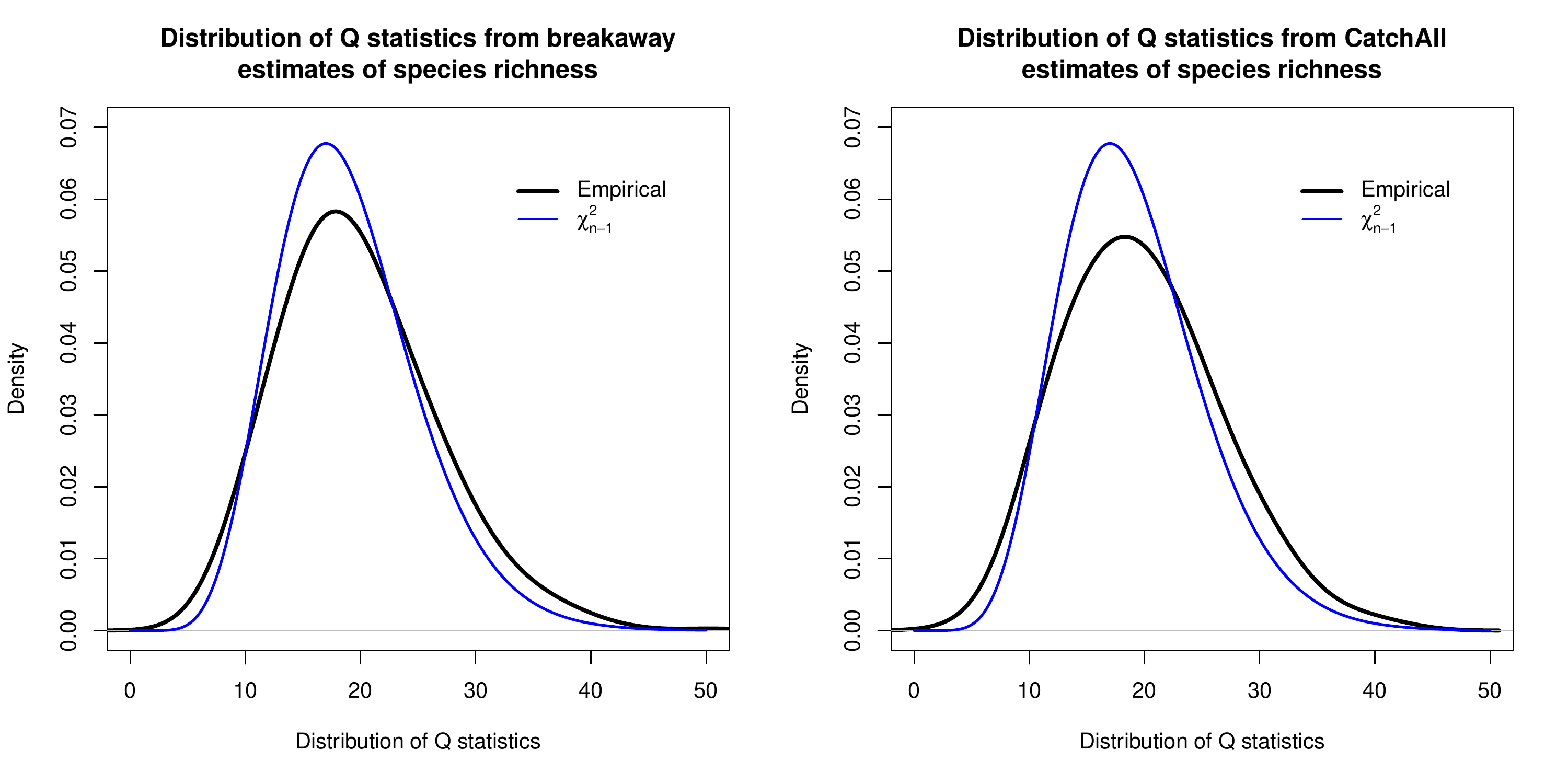}
\caption{Richness estimates derived from software breakaway (left) and CatchAll (right) appear to be approximately normally distributed with correct standard error in the negative binomial counts case.}
\label{fig2}  
\end{figure}
\end{center}

The model described above involves two major modelling assumptions: the estimate $\hat{C}_i$ is approximately normally distributed around $C_i$, and that the standard error in the estimate of $C_i$ provides a good approximation to the true standard deviation of $\hat{C}_i$ around $C_i$. Unfortunately, it is rare to know the true species richness for a high diversity situation, and so we proceed to investigate these assumptions via simulation. The negative binomial distribution has a long history in modelling frequency data, and while it rarely captures the large number of rare species that are present in many studies of practical interest, it provides a tractable option for model testing. In order to best reflect the large-unobserved diversity case, we draw 5,000 samples from a negative binomial distribution with probability parameter 0.99 and size parameter 500. We then omit the zeros from our samples (these represent the unobserved species) and attempt to estimate the number of species that we removed based only on the observed species. Repeating this procedure gives us a sequence $\hat{C}_1,\ldots,\hat{C}_m$ and $\hat{\sigma}_1,\ldots,\hat{\sigma}_m$. We may test the assumption of normality of the estimates around $C$ with approximately correct standard errors by observing the distribution of $\frac{\hat{C}_i - C}{\hat{\sigma}_i}$ for $C=5000$ and comparing it with a $\mathcal{N}(0,1)$ distribution. We compare the results of this procedure under two methods of estimating $\hat{C}$: breakaway, a ratio-based method developed for the high-diversity case \cite{rbmeunc}, and CatchAll, a computationally intensive method that fits a large number of mixed-Poisson models and selects the model of best fit \cite{catchall}. Figure \ref{fig1} compares the empirical distribution with a standard normal distribution. The sample mean of the breakaway rescaled estimates was $-5.11 \times 10^{-4}$ and the standard deviation was $1.0229$. The equivalent values for CatchAll were $-0.1459$ and $1.0202$. We note only slight deviation from the $\mathcal{N}(0,1)$ curve in each case and conclude that the assumption of normality is plausible, at least for frequency counts that follow an approximate negative binomial distribution. 
While the assumption of normality appears to hold, the effect of the substitution of $\hat{\sigma}_i$ for $\sigma_i$ may affect the test of homogeneity, and so we investigate the Type 1 error rate for this test. Similar to the above, we mimic the data-generating process of homogeneous negative binomial draws, but now observe the distribution of the homogeneity test statistics, $Q$, and compare their empirical distribution to a $\chi^2_{n-1}$ distribution. We draw 10,000 samples and partition them into 500 samples of size 20 (20 ``replicates''), compute the richness estimates and standard errors for each sample, fit an intercept-only model for each sample, and observe the distribution of the $Q$ statistics. Our theory suggests that the $Q$ statistics should be approximately distributed according to a $\chi^2_{19}$ distribution. We observe from Figure \ref{fig2} that the effect of the substitution $\hat{\sigma}_i$ leads to heavier distribution tails than would be expected under a $\chi^2_{19}$ distribution. The empirical Type 1 error rate for breakaway estimates is 7.2\%, and for CatchAll the error rate is 7.1\%. We conclude that substituting $\hat{\sigma}_i$ for $\sigma_i$ leads to a slightly exaggerated probability of rejecting a correct hypothesis of homogeneity and suggest caution when a computed test statistic is only marginally significant.

\subsection*{Case study: Gut microbiota richness is affected by antibiotics}
We now demonstrate the applicability of the above methodology in determining the effect of antibiotics on the gut microbiome. Dethlefsen {\it et. al} \cite{gutmicrobiome} employed pyrosequencing technology to obtain more than 7,000 rRNA sequences for three human subjects before, during, and after a course of ciprofloxacin. They observed that the treatment lead to an overall decrease in the observed richness of the microbiota communities but were unable to test this formally. We formally investigate this using their dataset.

\begin{center}
\begin{figure}
\includegraphics[trim= 0mm 60 10 40, scale=1]{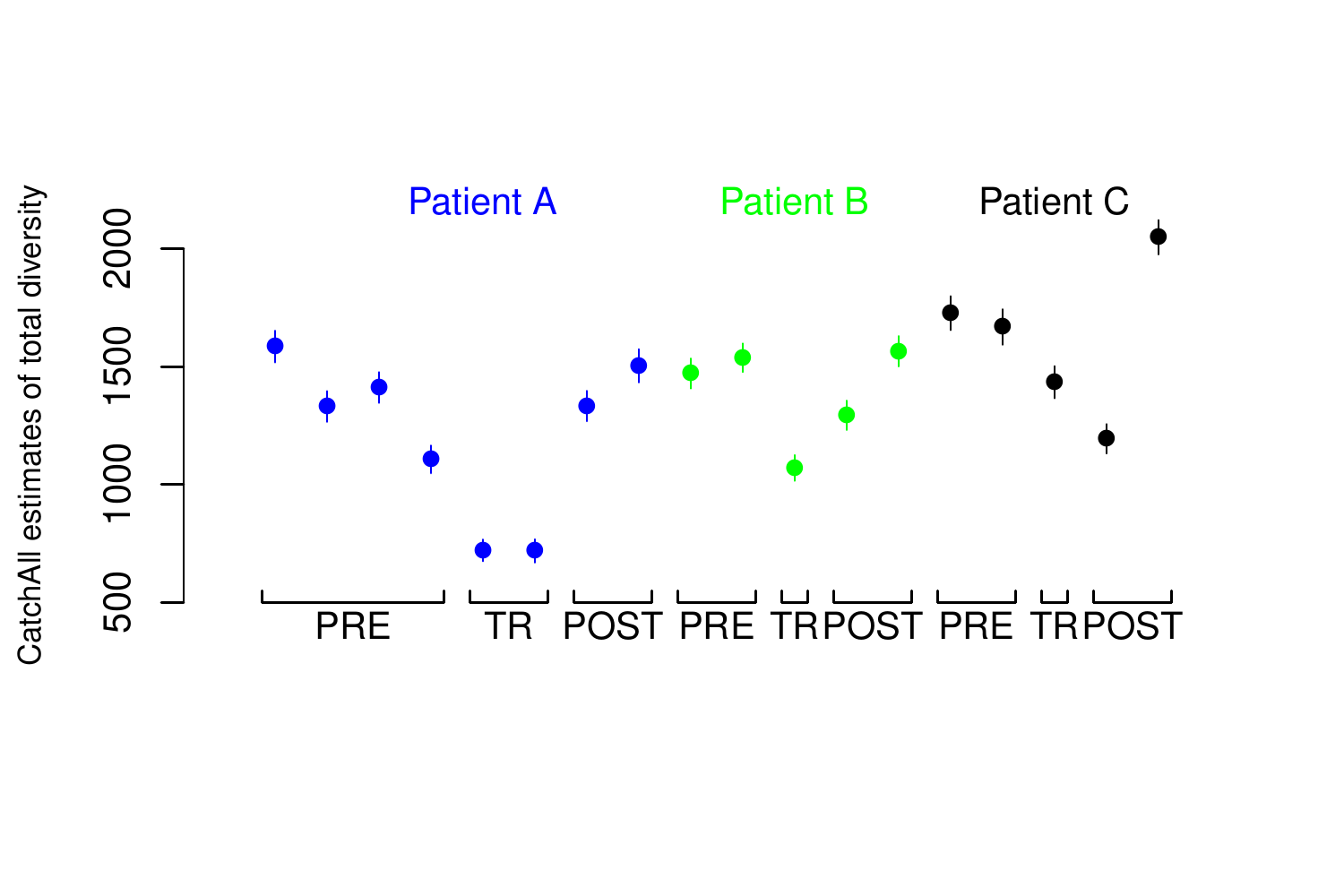}
\caption{CatchAll richness estimates for the Dethlefsen {\it et. al} dataset and 95\% Wald-type confidence intervals. Estimates were computed for each of three patients and pre-treatment (PRE), during treatment (TR) and post-treatment (POST).} 
\label{fig3}  
\end{figure}
\end{center}

A brief perusal of the frequency count tables suggests that we are in the medium diversity setting, and for this reason we use CatchAll \cite{catchall} to estimate species richness for each individual and each sample. We fit a model with treatment (pre-treatment, during treatment and post-treatment) and patient (Patient A, Patient B and Patient C) as predictors, referring to Figure \ref{fig3} to justify our choice of an additive model. We conclude that the model is highly significant in explaining richness ($Q=686.84, p<0.001$). In particular, we find that treatment is highly significant in decreasing richness ($p<0.001$), reducing diversity by 456 species on average. However, we find that there is no significant post-treatment affect ($p=0.638$) and that diversity recovers to pre-treatment levels after 4 weeks. These results concur with the general conclusions of Dethlefsen {\it et. al} \cite{gutmicrobiome}, but we emphasize that this methodology provides a formal approach to testing their hypothesis.

\subsection*{Case study: Biological replicates of soil samples are homogeneous}

Soil microbial communities are perhaps the most species-rich of all studied environments on Earth \cite{file}. Housing complex interfaces between the hydrosphere, atmosphere, lithosphere, and biosphere, soils exhibit extreme microscale heterogeneity in potential microbial habitats \cite{nwycr, trgk}, which may support the persistence of microbial species diversity \cite{lokn}. The complexity of these communities pose considerable challenges for diversity analysis and thus provide an extremely interesting test case. 

Whitman {\it et al.} \cite{theas} extracted, amplified, and sequenced bacterial 16S DNA with soils from a field trial with no amendments, with pyrogenic organic matter additions, and with fresh biomass additions. Treatments were laid out using a spatially balanced complete block design as described in \cite{vgsv}. Sampling took place less than 24 hours after additions, after 12 days, and after 82 days, with two samples taken less than 15 cm apart and combined for each plot. 

\begin{center}
\begin{figure}
\includegraphics[trim= 0mm 10 10 40, scale=0.8]{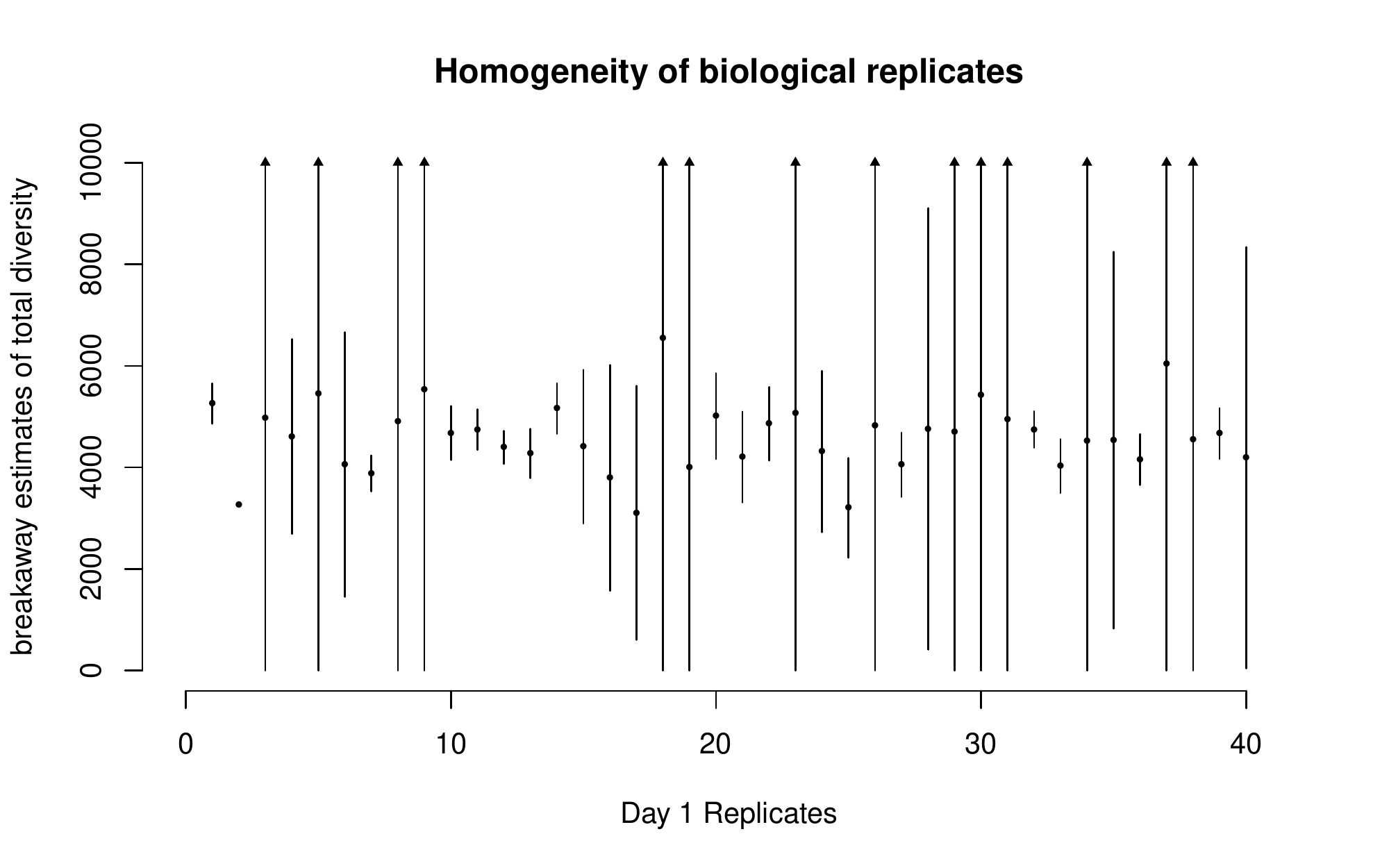}
\caption{CatchAll richness estimates for the Dethlefsen {\it et. al} dataset and 95\% Wald-type confidence intervals. Estimates were computed for each of three patients and pre-treatment (PRE), during treatment (TR) and post-treatment (POST).} 
\label{fig3a}  
\end{figure}
\end{center}

We focus on homogeneity of the biological replicates. All Day 1 samples may be considered to be replicates regardless of amendment, because the amendments were not yet incorporated. Symmetric confidence intervals for breakaway estimates of species richnesses for Day 1 samples may be seen in Figure \ref{fig3a}. An intercept-only model for the species richnesses initially rejects the null hypothesis for homogeneity ($p<0.0001$). However, noting that  the confidence interval for richness in Sample 2 appears to be unrealistically tight and potentially exerting high influence on the model, we re-test for homogeneity with this sample excluded to find that the test for heterogeneity is only marginally significant ($p=0.0296$). Given the conservative effect of the substitution $\hat{\sigma}_i$ discussed above, we do not reject the null hypothesis and conclude that the Day 1 replicates are in fact homogenous with respect to richness. 

\section*{Discussion}

The model for species richness presented above has many advantages compared to the popular approach of fitting a linear model for observed species richness with covariates. The most important development presented here is a model that accounts for unobserved diversity, which is usually ignored in formal diversity comparisons. 

While one of the advantages of the approach is its simplicity as a natural extension of the linear model, it shares a number of the limitations of the linear model. As a supervised procedure, the practitioner must make sensible choices of predictor variables that balance explanatory power with parsimony. Outliers (richness estimates with small standard errors) may exert influence on the model and visual diagnostics should be employed to determine the appropriateness of the model. Failing to confirm model assumptions and to check for outliers may lead to unreliable tests for significance, or false conclusions of heterogeneity.
Furthermore, richness estimation is a prediction problem, and so it is not possible to implement informative nonparametric approaches \cite{mao}. For this reason, estimation of species richness requires modelling assumptions. All known models suffer from estimator instability, or implausibility of model fit \cite{rbmeunc,bwwa}, which must be traded off when choosing an estimator of species richness. There is no guarantee that conclusions based on the mixed effects model described above are robust to choices of richness estimator, and we advise caution in this regard. 


It has previously been noted in the literature that observed diversity is highly sensitive to sampling depth (referred to as {\it sequencing depth} in the microbial case). This has lead to the common practice of {\it rarefying} the data to achieve equal sample sizes for each sample. For an excellent discussion of why rarefying should be outlawed, see McMurdie and Holmes \cite{wnwn}. Note that our method proposes modelling total diversity, not observed diversity. Total diversity is a fixed but unknown parameter, and is hence invariant to sampling depth. Estimated total diversity is a function of the sample, but it is more robust to sampling depth than observed diversity. While we advocate additional sampling and sequencing whenever possible, we note that our method accounts for inexhaustive sampling and emphasize that abundances should not be rarefied prior to implementing the methodology.

We believe that the greatest advantage of this method is the ability to  {\it share strength} across samples to improve inference via the BLUP framework of the linear mixed model. Large numbers of rare taxa may prohibit accurate and precise estimation of total diversity from single samples, but pooling samples believed to have similar structure can vastly reduce standard errors in estimation of total diversity for each of the observed samples, even if their covariate information differs. We believe that this is the first proposal to use community composition information (``beta diversity'') to improve inference about individual samples (``alpha diversity'').

\section*{Conclusions}

We present the first approach to modelling total species richness as a function of covariates. Existing methods fail to account for unobserved diversity by only noting changes in observed diversity
. We believe the strength of the approach is most pronounced in high-diversity environments, where unobserved diversity may dominate. We demonstrate the approach on two microbial datasets: a soil dataset where we conclude homogeneity of biological replicates, and a gut microbiome dataset where we conclude that an antibiotic significantly decreases taxonomic richness. We believe that the method is highly general and applicable to any situation where community abundance data is available, including microbial and animal abundances. The relevant software for diversity estimation and prediction is given in the R package ``breakaway'' \cite{breakaway}, with the methodology accessible via the function ``betta''. Given the importance of maintaining stable ecological diversity \cite{vaginalmicrobiome} and  ongoing investigation of data obtained by the Human Microbiome Project \cite{hmp1,hmp2}, we hope that the inferential procedure outlined here will inform local and global policy, especially with respect to human health (for example, via gut microflora stabilization) and climate change (for example, via carbon fixing soil treatments).

\bibliography{betta}

\end{document}